\documentclass[%
  reprint,
  superscriptaddress,
  nofootinbib,
  amsmath,amssymb,
  aps,
  prc,
]{revtex4-1}

\usepackage{graphics}
\usepackage{graphicx}
\usepackage{dcolumn}
\usepackage{bm}

\usepackage{color,hyperref}
\hypersetup{colorlinks,breaklinks,
  linkcolor=blue,urlcolor=blue,
  anchorcolor=blue,citecolor=blue}

\usepackage[mathlines]{lineno}
\usepackage{lipsum}
\usepackage{tikz}
\usetikzlibrary{shapes.geometric,decorations.fractals,shadows,shapes.multipart,arrows,snakes,positioning,arrows}
\usetikzlibrary{decorations.pathmorphing,decorations.markings}
\usetikzlibrary{calc,intersections}
\usepackage{comment}
\usepackage{textcomp}
\usepackage{bigints}
\usepackage{enumitem}

\newcommand{\onbb}{$0\nu\beta\beta$}
\newcommand{\bec}{$0\nu\beta^{\!^+}\hspace{-0.2em}EC$}
\newcommand{\onpp}{$0\nu\beta^{\!^+}\hspace{-0.4em}\beta^{\!^+}$\hspace{-0.2em}}
\newcommand{\onecec}{$0\nu ECEC$}
\newcommand{\p}{$\beta^{\!^+}$\hspace{-0.2em}}

\newcommand{\Te}{$^{130}$Te}
\newcommand{\Tee}{$^{120}$Te}

\newcommand{\cuore}{CUORE}
\newcommand{\cuoreo}{\mbox{CUORE-0}}
\newcommand{\cuoricino}{Cuoricino}
\newcommand{\TeOO}{TeO$_2$}
\newcommand{\bp}{$\beta^{\!^+}$\hspace{-0.2em}}

\definecolor{arsenic}{rgb}{0.23,0.27,0.29}
\definecolor{grey}{rgb}{0.6,0.6,0.6}
\definecolor{darkblue}{RGB}{0,0,127}
\definecolor{darkgreen}{RGB}{0,170,0}
\definecolor{darkred}{RGB}{220,0,0}
\definecolor{lightblue}{RGB}{150,200,255}

\begin{document}


\title{Search for Neutrinoless \boldmath$\beta^{\!^+}\hspace{-0.2em}EC$ Decay
  of \boldmath\Tee\ with \cuoreo}


\author{C.~Alduino}
\affiliation{Department of Physics and Astronomy, University of South Carolina, Columbia, SC 29208, USA}

\author{K.~Alfonso}
\affiliation{Department of Physics and Astronomy, University of California, Los Angeles, CA 90095, USA}

\author{D.~R.~Artusa}
\affiliation{Department of Physics and Astronomy, University of South Carolina, Columbia, SC 29208, USA}
\affiliation{INFN -- Laboratori Nazionali del Gran Sasso, Assergi (L'Aquila) I-67100, Italy}

\author{F.~T.~Avignone~III}
\affiliation{Department of Physics and Astronomy, University of South Carolina, Columbia, SC 29208, USA}

\author{O.~Azzolini}
\affiliation{INFN -- Laboratori Nazionali di Legnaro, Legnaro (Padova) I-35020, Italy}

\author{G.~Bari}
\affiliation{INFN -- Sezione di Bologna, Bologna I-40127, Italy}

\author{F.~Bellini}
\affiliation{Dipartimento di Fisica, Sapienza Universit\`{a} di Roma, Roma I-00185, Italy}
\affiliation{INFN -- Sezione di Roma, Roma I-00185, Italy}

\author{G.~Benato}
\affiliation{Department of Physics, University of California, Berkeley, CA 94720, USA}

\author{A.~Bersani}
\affiliation{INFN -- Sezione di Genova, Genova I-16146, Italy}

\author{M.~Biassoni}
\affiliation{Dipartimento di Fisica, Universit\`{a} di Milano-Bicocca, Milano I-20126, Italy}
\affiliation{INFN -- Sezione di Milano Bicocca, Milano I-20126, Italy}

% ...

\collaboration{\cuore\ Collaboration}

\date{\today}

\begin{abstract}
  \ \\
  \noindent   We have performed a search for neutrinoless $\beta^{\!^+}\hspace{-0.2em}EC$ decay of \Tee\
  using the final \cuoreo\ data release.
  We describe a new analysis method for the simultaneous fit of signatures
  with different event topology, and of data subsets with different 
  signal efficiency, obtaining a limit on the half-life of the decay of
  $T_{1/2}>1.6\cdot10^{21}$~yr at $90\%$~CI.
  Combining this with results from \cuoricino, a predecessor experiment,
  we obtain the strongest limit to date, corresponding to $T_{1/2}>2.7\cdot10^{21}$~yr at $90\%$~CI.
  
\end{abstract}

\pacs{Valid PACS appear here}
\maketitle

\section{Introduction}\label{sec:Introduction}

The search for neutrinoless double beta (\onbb) decay~\cite{Furry:1939qr,DellOro:2016tmg}
aims at answering questions regarding
the conservation of total lepton number~\cite{Fukugita:1986hr},
the Majorana or Dirac nature of neutrinos~\cite{Vergados:2012xy}, and
the mechanism inducing non-zero splittings between the neutrino mass
eigenvalues~\cite{PDG,Kajita:2016cak,McDonald:2016ixn,Eguchi:2002dm,An:2016ses,Seo:2016uom}.
The process can consist of the emission of two electrons -- which is the most commonly
investigated option -- or of two positrons (\onpp). In the latter case, one or both positrons
can be substituted by an electron capture ($EC$).
For \onecec\ decay, the decay rate is typically suppressed because an additional radiative process is required
by energy and momentum conservation. Hence, the \onpp\ and \bec\ decays are more interesting
from the experimental perspective.

The Cryogenic Underground Observatory for Rare Events (\cuore)~\cite{CUORE-proposal,Artusa:2014lgv}
and \cuoreo~~\cite{Aguirre:2014lua,Alduino:2016vjd}
are experiments searching for the \onbb\ decay of \Te\ with \TeOO\ crystals operated as bolometric detectors.
The use of tellurium with natural isotopic composition in the crystals
also allows us to search for the decay of isotopes other than \Te.
In particular, \Tee, present with a natural abundance of $0.09(1)\%$~\cite{iupac2016},
can decay via \onecec\ and via \bec\ disintegration.
In this work, we present the search for \bec\ decay using \cuoreo\ data.
At present, no calculation of the nuclear matrix element is available in literature for \Tee\ decay.
For other isotopes, the expected \bec\ half-lives are a few orders of magnitude larger
than that of \onbb\ decay for the most commonly investigated cases~\cite{Barea:2013wga,Kotila:2014zya}.
Despite this, and the low abundance of \Tee,
the presence of the \p\ with the consequent emission of a pair
of back-to-back 511~keV $\gamma$ rays provides extremely clean signatures of \bec\ decays.
The \bec\ decay of \Tee\ can be written as:
\begin{align*}
  ^{120}\text{Te} + e_b^{-} & \rightarrow ^{120}\text{Sn}^* + \beta^{\!^+} \\
  & \rightarrow ^{120}\text{Sn} + X + \beta^{\!^+} \\
  & \rightarrow ^{120}\text{Sn} + X + 2\gamma_{511}\ ,
\end{align*}
where $e_b^-$ indicates the atomic electron captured from a shell with binding energy $E_b$,
while $X$ indicates an Auger electron or an X-ray emitted in the process.
In \bec\ decay the available energy is shared between the four emitted particles,
with the daughter nucleus being almost at rest because of its larger mass.
Here, we assume that the X-ray or the Auger electron are fully absorbed
in the same crystal where the decay takes place.
Namely, the K-shell  binding energy of tin is $29.2$~keV~\cite{Bearden:1967gqa}.
Electrons and $\gamma$ rays of this energy have a chance to escape the crystal only if they are emitted
at $\lesssim10$~{\textmu}m from the surface, thus only the decays taking place
in a negligible fraction of the crystal volume would be affected by energy loss.

Given the absence of neutrinos carrying away part of the available energy,
the kinetic energy $K$ of the emitted positron is peaked at $K = Q - 2m_e - E_b$,
where $Q$ is the Q-value of the reaction:
\begin{equation}
  Q = m(^{120}\text{Te})-m(^{120}\text{Sn})\ ,
\end{equation}
and $m$ are the masses of the considered nuclei.
Only one direct measurement of $Q$, obtained with a Penning trap, is available in literature,
i.e. $Q=1714.8\pm1.3$~keV~\cite{Scielzo:2009nh}.

\begin{figure}[]

  \begin{tikzpicture}[xscale=1,yscale=1]

    \definecolor{lgrey}{rgb} {0.88,0.88,0.88}
    
    \def\minX{0.8}
    \def\deltaX{0.1}
    \def\minY{0.2}
    \def\deltaY{0.5}
    \def\Delta{1}
    \def\thrleft{\Delta+\deltaX}

    \def\betaXleft{\minX+0.8*\Delta}

    \draw[fill=lgrey] (\minX,               \minY+\thrleft) rectangle (\minX+\Delta,          \minY+\Delta+\thrleft);
    \draw (\minX+\Delta+\deltaX,\minY+\thrleft) rectangle (\minX+2*\Delta+\deltaX,\minY+\Delta+\thrleft);
    \node at (0,\minY+\thrleft+0.5*\Delta) (c){(2)};
    \draw (c);
    
    \node[star,fill=red,scale=.7,star point ratio=0.4] at (\minX+0.5\Delta,\minY+\thrleft+0.5*\Delta) (c1){};
    \node at (\minX+\Delta,\minY+\thrleft) (c2){};
    \node at (\minX,\minY+\thrleft+\Delta) (c3){};
    \draw (c1);
    \draw[snake=coil, line after snake=.5mm, segment aspect=0,%
      segment length=5pt,segment amplitude=1,-stealth] (c1) -- (c2);
    \draw[snake=coil, line after snake=.5mm, segment aspect=0,%
      segment length=5pt,segment amplitude=1,-stealth] (c1) -- (c3);

    \draw[fill=lgrey] (\minX,               \minY+\Delta+\deltaY+2*\thrleft) rectangle (\minX+\Delta,\minY+2*\Delta+\deltaY+2*\thrleft);
    \draw (\minX+\Delta+\deltaX,\minY+\Delta+\deltaY+2*\thrleft) rectangle (\minX+2*\Delta+\deltaX,\minY+2*\Delta+\deltaY+2*\thrleft);
    \node at (0,\minY+1.5\Delta+\deltaY+2*\thrleft) (b){(1)};
    \draw (b);

    \node[star,fill=red,scale=.7,star point ratio=0.4] at (\minX+0.85\Delta,\minY+1.5*\Delta+\deltaY+2*\thrleft) (b1){};
    \node at (\minX,\minY+1.5*\Delta+\deltaY+2*\thrleft) (b2){};
    \node at (\minX+2.5*\Delta+\deltaX,\minY+1.5*\Delta+\deltaY+2*\thrleft) (b3){};
    \draw (b1);
    \draw[snake=coil, line after snake=.5mm, segment aspect=0,%
      segment length=5pt,segment amplitude=1,-stealth] (b1) -- (b2);
    \draw[snake=coil, line after snake=.5mm, segment aspect=0,%
      segment length=5pt,segment amplitude=1,-stealth] (b1) -- (b3);
    
    \draw[fill=lgrey] (\minX,\minY+2*\Delta+2*\deltaY+2*\thrleft) rectangle (\minX+\Delta,\minY+3*\Delta+2*\deltaY+2*\thrleft);
    \draw (\minX+\Delta+\deltaX,\minY+2*\Delta+2*\deltaY+2*\thrleft) rectangle (\minX+2*\Delta+\deltaX,\minY+3*\Delta+2*\deltaY+2*\thrleft);
    \node at (0,\minY+2.5*\Delta+2*\deltaY+2*\thrleft) (a){(0)};
    \draw (a);

    \node[star,fill=red,scale=.7,star point ratio=0.4] at (\minX+0.85\Delta,\minY+2.5*\Delta+2*\deltaY+2*\thrleft) (a1){};
    \node at (\minX-0.5*\Delta,\minY+2.5*\Delta+2*\deltaY+2*\thrleft) (a2){};
    \node at (\minX+2.5*\Delta+\deltaX,\minY+2.5*\Delta+2*\deltaY+2*\thrleft) (a3){};
    \draw (a1);
    \draw[snake=coil, line after snake=.5mm, segment aspect=0,%
      segment length=5pt,segment amplitude=1,-stealth] (a1) -- (a2);
    \draw[snake=coil, line after snake=.5mm, segment aspect=0,%
      segment length=5pt,segment amplitude=1,-stealth] (a1) -- (a3);

    \def\gap{2}
    \draw[fill=lgrey] (\minX+2*\Delta+\deltaX+\gap,  \minY) rectangle (\minX+3*\Delta+\deltaX+\gap,  \minY+\Delta);
    \draw (\minX+3*\Delta+2*\deltaX+\gap,\minY) rectangle (\minX+4*\Delta+2*\deltaX+\gap,\minY+\Delta);
    \draw[fill=lgrey] (\minX+2*\Delta+\deltaX+\gap,  \minY+\Delta+\deltaX) rectangle (\minX+3*\Delta+\deltaX+\gap,  \minY+2*\Delta+\deltaX);
    \draw (\minX+3*\Delta+2*\deltaX+\gap,\minY+\Delta+\deltaX) rectangle (\minX+4*\Delta+2*\deltaX+\gap,\minY+2*\Delta+\deltaX);
    \draw[fill=lgrey] (\minX+2*\Delta+\deltaX+\gap,  \minY+2*\Delta+2*\deltaX) rectangle (\minX+3*\Delta+\deltaX+\gap,  \minY+3*\Delta+2*\deltaX);
    \draw (\minX+3*\Delta+2*\deltaX+\gap,\minY+2*\Delta+2*\deltaX) rectangle (\minX+4*\Delta+2*\deltaX+\gap,\minY+3*\Delta+2*\deltaX);
    \node at (\minX+2*\Delta+\deltaX+0.65*\gap,\minY+\thrleft+0.5*\Delta) (f){(5)};
    \draw (f);

    \node[star,fill=red,scale=.7,star point ratio=0.4] at (\minX+2.5\Delta+\deltaX+\gap,\minY+0.5*\Delta+\thrleft) (f1){};
    \node at (\minX+2.5*\Delta+\deltaX+\gap,\minY+0.4*\Delta) (f2){};
    \node at (\minX+2.5*\Delta+\deltaX+\gap,\minY+1.7*\Delta+\thrleft) (f3){};
    \draw (f1);
    \draw[snake=coil, line after snake=.5mm, segment aspect=0,%
      segment length=5pt,segment amplitude=1,-stealth] (f1) -- (f2);
    \draw[snake=coil, line after snake=.5mm, segment aspect=0,%
      segment length=5pt,segment amplitude=1,-stealth] (f1) -- (f3);
    
    \draw[fill=lgrey] (\minX+2*\Delta+\deltaX+\gap,  \minY+\Delta+\deltaY+2*\thrleft) rectangle (\minX+3*\Delta+\deltaX+\gap,  \minY+2*\Delta+\deltaY+2*\thrleft);
    \draw[fill=lgrey] (\minX+3*\Delta+2*\deltaX+\gap,\minY+\Delta+\deltaY+2*\thrleft) rectangle (\minX+4*\Delta+2*\deltaX+\gap,\minY+2*\Delta+\deltaY+2*\thrleft);
    \node at (\minX+2*\Delta+\deltaX+0.65*\gap,\minY+1.5\Delta+\deltaY+2*\thrleft) (e){(4)};
    \draw (e);

    \node[star,fill=red,scale=.7,star point ratio=0.4] at (\minX+2.85\Delta+\deltaX+\gap,\minY+1.5*\Delta+\deltaY+2*\thrleft) (e1){};
    \node at (\minX+2*\Delta+\deltaX+\gap,\minY+1.5*\Delta+\deltaY+2*\thrleft) (e2){};
    \node at (\minX+3.85*\Delta+\deltaX+\gap,\minY+1.5*\Delta+\deltaY+2*\thrleft) (e3){};
    \draw (e1);
    \draw[snake=coil, line after snake=.5mm, segment aspect=0,%
      segment length=5pt,segment amplitude=1,-stealth] (e1) -- (e2);
    \draw[snake=coil, line after snake=.5mm, segment aspect=0,%
      segment length=5pt,segment amplitude=1,-stealth] (e1) -- (e3);
    
    \draw[fill=lgrey] (\minX+2*\Delta+\deltaX+\gap,  \minY+2*\Delta+2*\deltaY+2*\thrleft) rectangle (\minX+3*\Delta+\deltaX+\gap,  \minY+3*\Delta+2*\deltaY+2*\thrleft);
    \draw[fill=lgrey] (\minX+3*\Delta+2*\deltaX+\gap,\minY+2*\Delta+2*\deltaY+2*\thrleft) rectangle (\minX+4*\Delta+2*\deltaX+\gap,\minY+3*\Delta+2*\deltaY+2*\thrleft);
    \node at (\minX+2*\Delta+\deltaX+0.65*\gap,\minY+2.5*\Delta+2*\deltaY+2*\thrleft) (d){(3)};
    \draw (d);

    \node[star,fill=red,scale=.7,star point ratio=0.4] at (\minX+2.85\Delta+\deltaX+\gap,\minY+2.5*\Delta+2*\deltaY+2*\thrleft) (d1){};
    \node at (\minX+1.5*\Delta+\deltaX+\gap,\minY+2.5*\Delta+2*\deltaY+2*\thrleft) (d2){};
    \node at (\minX+3.85*\Delta+\deltaX+\gap,\minY+2.5*\Delta+2*\deltaY+2*\thrleft) (d3){};
    \draw (d1);
    \draw[snake=coil, line after snake=.5mm, segment aspect=0,%
      segment length=5pt,segment amplitude=1,-stealth] (d1) -- (d2);
    \draw[snake=coil, line after snake=.5mm, segment aspect=0,%
      segment length=5pt,segment amplitude=1,-stealth] (d1) -- (d3);
    
  \end{tikzpicture}
  \caption{\label{fig:signatures}Signatures of \Tee\ \bec\ decay in \cuoreo.
    The red stars represent the $\beta^+$ energy depositions,
    and the arrows represent the $511$~keV $\gamma$ rays following its annihilation.
    The shaded squares represent the crystals with non-zero energy depositions.
    We do not apply any distance cut for signatures (3)--(5) and accept also events
    depositing energy in non-neighboring crystals.}
\end{figure}
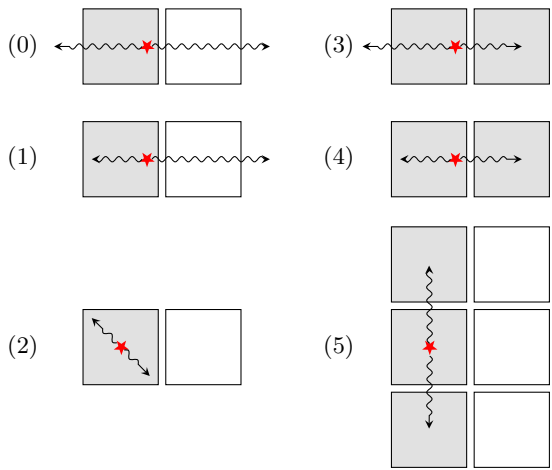

The energy deposited inside the crystal where the decay takes place,
$\mu$, is the sum of the kinetic energy of the positron
and that of the X-ray or Auger electron.
Since, in most of the cases, the positron is fully absorbed, we can write:
\begin{equation}
  \mu = K + E_b = Q - 2 m_e = 692.8\pm1.3~\text{keV}\ .
\end{equation}
The expected energy spectrum for \bec\ decay in the detector where the process occurs
is therefore a peak at $\mu$. 
If we also consider the two $511$~keV $\gamma$ rays, six different signatures are possible.
These are  depicted in Fig.~\ref{fig:signatures} and reported in Table~\ref{tab:signatures}
(we refer them with the symbol $(s)$, with $s=0,\dots,5$).
Each $\gamma$ can either be absorbed in the same crystal, or in a different crystal,
or escape the detector volume and be absorbed elsewhere.

The most stringent limit on \Tee\ \bec\ half-life, \mbox{$T_{1/2}>1.9\cdot10^{21}$~yr} at $90\%$~confidence level
(C.L.)~\cite{Andreotti:2010nn}, was obtained by \cuoricino~\cite{Arnaboldi:2004qy}.
The exposure of \cuoreo\ is just about half of the \cuoricino\ one,
however the lower background and higher signal efficiency lead to a higher sensitivity.
Moreover, in the present work we develop an analysis method which fully exploits the information
available in all the six different \bec\ decay signatures.
These factors allow to reach a sensitivity to \bec\ decay comparable to that of \cuoricino.
\cuore\ will have a much higher sensitivity due primarily
to the larger mass and efficiency of detecting the two 511~keV gammas.

\begin{table*}[]
  \caption{Signatures of \Tee\ \bec\ decay in \cuoreo. 
    For each signature ($s$) we report the signal peak position $\vec{\mu}$ of \bec\ decay,
    the multiplicity $M$ corresponding to the number of crystals with a non-zero energy deposition,
    and the fit range(s) $\Delta E_i$, with $i=1,\dots,M$. The range of signature $(0)$ is narrower
    than all others due to the presence of shoulders at $\sim640$~keV and $\sim740$~keV which would
    require the parameterization of the continuum background with a high-order polynomial.
    The last two columns report the average containment efficiency $\varepsilon_{\rm MC}$, i.e. the probability
    of having a full energy deposition in the detector volume. We obtain this from MC simulations,
    and report it with its relative binomial uncertainty for floors 2--12 and 1,13.
  }
  \label{tab:signatures}
  \begin{ruledtabular}
    \begin{tabular}{llcccccc}
      &                   &     & \multicolumn{3}{c}{Energy range [keV]} & \multicolumn{2}{c}{$\varepsilon_{\rm MC}[\%]$} \\
      Signature & $\vec{\mu}$~[keV] & $M$ & $\Delta E_1$ & $\Delta E_2$ & $\Delta E_3$ &Fl. $2$ -- $12$ & Fl. $1,13$ \\
      \hline\\ [-2ex]
      (0) & $692.8$           & 1 & $[657,720]$   &             &             & $0.162(2)$ & $0.309(4)$  \\
      (1) & $1203.8$          & 1 & $[1150,1250]$ &             &             & $1.23(2)$ & $1.60(2)$  \\
      (2) & $1714.8$          & 1 & $[1665,1775]$ &             &             & $0.90(1)$ & $0.92(1)$  \\
      (3) & $(692.8,511)$     & 2 & $[650,750]$   & $[460,560]$ &             & $0.317(3)$ & $0.303(3)$  \\
      (4) & $(1203.8,511)$    & 2 & $[1150,1250]$ & $[460,560]$ &             & $0.657(5)$ & $0.471(4)$   \\
      (5) & $(692.8,511,511)$ & 3 & $[650,750]$   & $[460,560]$ & $[460,560]$ & $0.0559(5)$ & $0.0196(3)$ \\
    \end{tabular}
  \end{ruledtabular}
\end{table*}

\section{\cuoreo}\label{sec:cuore0}

\cuoreo\ was a prototype of \cuore\ operated  between 2013 and 2015.
In addition to being a test stand for the \cuore\ assembly~\cite{Buccheri:2014bma}
and cleaning procedures~\cite{Alessandria:2011vj,Alessandria:2012zp},
\cuoreo\ provided data leading to competitive physics results~\cite{Alfonso:2015wka,Alduino:2016vtd}.
\cuoreo\ consisted of 52~\TeOO\ crystals with natural Te composition operated as source
and detector for the \onbb\ decay of \Te.
The crystals are 5~cm cubes mounted in a tower of 13~floors, with 4 crystals per floor.
They were operated as cryogenic calorimeters (bolometers) at a temperature of $\sim10$~mK
and read-out with neutron transmutation doped germanium thermistors.
The total \TeOO\ mass is $39$~kg. Using the most recent evaluation
of the \Tee\ natural abundance, $f_{120}=0.09(1)\%$~\cite{iupac2016},
the \Tee\ mass contained in \cuoreo\ is $28$~g, corresponding to $1.3\cdot10^{23}$ atoms of \Tee.
We note that this value of $f_{120}$ differs from the $0.096(2)\%$ used by \cuoricino~\cite{Andreotti:2010nn},
which was taken from Ref.~\cite{nucleardata}.

\cuoreo\ used the same cryostat and shielding as \cuoricino~\cite{Andreotti:2010vj,Arnaboldi:2008ds}.
The shielding consists of two external layers of low radioactivity lead for a total thickness of 20~cm
and a 1.2~cm internal layer of cold ancient Roman lead~\cite{romanlead}.
The cryostat thermal shields are made of electrolytic copper which provides
an additional layer of shielding ($\sim1.5$~cm),
and the whole cryostat is enclosed in a 10~cm layer of borated polyethylene shielding.
The front end electronics and the data acquisition were the same as for \cuoricino.
For a more detailed description, see Refs.~\cite{Aguirre:2014lua,Arnaboldi:2008ds,Arnaboldi:2004jj,Arnaboldi:2010zz}.

\section{Data Analysis}\label{sec:data}

We use the entire \cuoreo\ data set, which corresponds
to $35.2$~kg$\cdot$yr of \TeOO\ exposure. We use the same data processing and selection
as described in Ref.~\cite{Alduino:2016zrl}, except for anti-coincidence cut,
since we now select events with multiplicity $M$ (i.e. numbers of crystals with
a non-zero energy deposition)
and energy which satisfy the criteria reported in Table~\ref{tab:signatures}.

The application of the selection cuts introduces an efficiency term, $\varepsilon_{\rm cut}$,
which is common to all signatures.
This is the product of the trigger and reconstruction efficiency, $\varepsilon_{\rm trigger}$,
and of the pile-up and Pulse Shape Analysis (PSA) efficiency, $\varepsilon_{\rm PSA}$.
We use the same values reported in~\cite{Alduino:2016zrl}, i.e. $\varepsilon_{\rm trigger}=98.529\pm0.004\%$
and $\varepsilon_{\rm PSA}=93.7\pm0.7\%$. The product of the two yields $\varepsilon_{\rm cut}=92.3\pm0.7\%$.
We apply these cuts independently to each channel,
exponentiating the efficiency term to the corresponding multiplicity: $\varepsilon_{\rm cut}^M$.

Additionally, the selection of events with $M=1$, $2$ or $3$ introduces
a further efficiency term, $\varepsilon_{\rm M}$.
We exploit the \cuoreo\ event rate ($\sim0.001$~Hz) to compute the probability of having random coincidences,
which induce pile-up events in the $M=1$, $M=2$ and $M=3$ spectra~\cite{spozzi},
obtaining the following coincidence efficiencies:
\mbox{$\varepsilon_{\rm M=1}=99(1)\%$}, \mbox{$\varepsilon_{\rm M=2}=99.2(1)\%$}
and \mbox{$\varepsilon_{\rm M=3}=98.8(1)\%$}.

Finally, we consider the containment efficiency, i.e. the probability for an event
of each signature to be fully contained in the \TeOO\ volume.
We compute the containment efficiency $\varepsilon_{\rm MC}$ using Monte Carlo (MC)
simulations (see Sec.~\ref{sec:mc}) and expect it to be floor dependent.
Specifically, in signatures $(0)$ and $(1)$ $\varepsilon_{\rm MC}$ should be larger
for the uppermost and lowermost floors (floors $1$ and $13$)
because these crystals only have neighbors on three sides rather than four,
hence the $\gamma$ rays have a higher chance of escaping undetected.
Instead, signature $(2)$, in which both $\gamma$ rays are absorbed in the same crystal
where the decay occurs, should have the same efficiency in all floors.
Finally, we expect signatures $(3)$, $(4)$ and $(5)$ to have a larger efficiency
for \bec\ decays taking place in the inner floors ($2$--$12$). 
Based on these considerations, we divide the data into subsets having the \bec\ decay
in floors $2$--$12$ (subset $0$) or floors $1$,$13$ (subset $1$).
We give more details on the computation of the containment efficiency in Sec.~\ref{sec:mc}.


We determine the energy resolution using the background peaks present in the \cuoreo\ $M=1$ physics spectrum,
and keeping the distinction between floors $1$, $13$ and $2$--$12$.
We fit the most prominent peaks in the energy spectrum:
the Annihilation Peak (AP) at $511$~keV and the Single Escape Peak (SEP),
plus a variety of $\gamma$ lines ranging from $238$~keV to $2615$~keV.
We find the AP and the SEP to be wider than the $\gamma$ lines both in calibration and physics data.
Signatures $(3)$, $(4)$ and $(5)$ feature the annihilation peak in the signal parametrization,
therefore we need to treat them separately from the $\gamma$ lines.
In signatures (1) and (4) a line at 1203.8~keV also appears in the signal parameterization:
this line corresponds to the sum energy of the \bp, the X-ray or Auger electron,
and a 511~keV annihilation $\gamma$, hence it is also expected to be broadened.

\begin{figure}[h]
  \centering
  \includegraphics[width=\columnwidth]{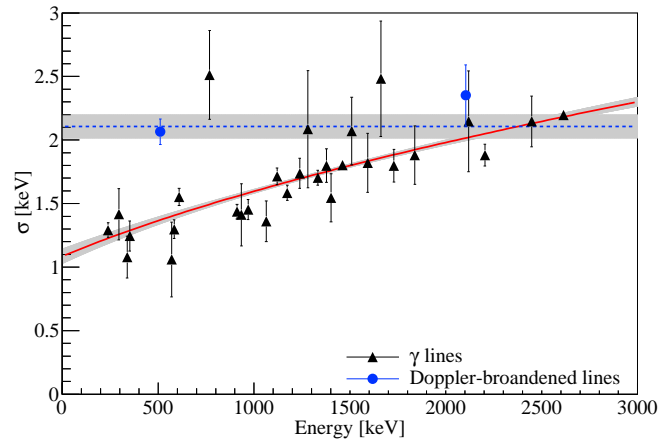}
  \caption{Resolution curves for $\gamma$ lines (black triangles, red solid curve) and broadened
    peaks (blue circles and blue dashed curve) for floors 2--12.}
  \label{fig:fwhm}
\end{figure}

We fit the energy resolution ($\sigma$) of the $\gamma$ lines in the \cuoreo\ physics spectrum
as a function of energy with the following function:
\begin{equation}\label{eq:fwhm}
  \sigma_{\gamma}(E) = \sqrt{a + b \cdot E}
\end{equation}
where $a$ describes the thermal and electronic noise, while $b$ is a parameter
connected to the phonon production and collection.
The fit results are reported in Table~\ref{tab:fwhm}.
On the other hand, the presence in the physics spectrum of only two broadened peaks prevents a proper fit
of energy resolution as a function of energy.
Therefore we take $\sigma_{\rm B}$, the resolution of the broadened lines,
to be the average of the resolutions of the AP and SEP (Table~\ref{tab:fwhm}).
The fits of $\sigma_{\gamma}$ and $\sigma_{\rm B}$ for floors 2--12 are also shown in Fig.~\ref{fig:fwhm}.
This procedure differs from that used in Ref.~\cite{Alfonso:2015wka}
in the separation between floors 1--13 and 2--12,
and in the ad-hoc treatment of the broadened lines.

\begin{table}[h]
  \caption{Best fit values for the parameters of Eq.~(\ref{eq:fwhm}),
    and $\sigma_{\rm B}$ of broadened peaks for subsets 0 and 1.
    The uncertainties correspond to the statistical errors of the fit.}
  \label{tab:fwhm}
  \begin{ruledtabular}
    \begin{tabular}{lccc}
      $d$ & $a$ [keV$^2$]  & $b$ [keV] & $\sigma_{\rm B}$ [keV] \\
      \hline\\ [-2ex]
      0 & 1.2(1) & $1.37(7)\cdot10^{-3}$ & 2.1(2) \\
      1 & 1.5(1) & $1.70(9)\cdot10^{-3}$ & 2.4(2) \\
    \end{tabular}
  \end{ruledtabular}
\end{table}

The energy resolution of the signal peaks (see Eq.~(\ref{eq:pdfsignal}) of Sec.~\ref{sec:statistics})
is taken as $\sigma_\gamma$ or $\sigma_{\rm B}$, depending on the presence
of broadening in the considered signal.
The same approach is used for the background components.

\section{Monte Carlo Simulations}\label{sec:mc}

We use MC simulations to extract the containment efficiency,
and to get an understanding of the most appropriate fit model for each signature.
Specifically, we use the background model described in Ref.~\cite{Alduino:2016vtd}
to define a maximal fit model that contains, for each signature and subset,
all components visible in the simulated spectra.

In order to compute the \Tee\ \bec\ decay containment efficiency,
we simulated $10^7$ positrons with $692.8$~keV kinetic energy uniformly distributed in the \TeOO\ volume.
We define the containment efficiency $\varepsilon_{\rm MC}$ for a given signature and subset
as the number of events which deposit energy in a $\pm4\sigma$ window around
the expected \bec\ decay peak position $\vec{\mu}$, divided by the number of generated primaries:
\begin{equation}\label{eq:effMC}
  \varepsilon_{\rm MC} = \frac{N(E \in [ \mu - 4 \sigma, \mu + 4 \sigma ] )}{ N_{tot}}\ .
\end{equation}
We chose the number of generated primaries in order to have $\sigma_{\varepsilon_{\rm MC}}/\varepsilon_{\rm MC} \sim 1\%$.
In all cases, we account for the non-operative channel in floor 10~\cite{Alfonso:2015wka}.
For the signatures with $M>1$ we also account for the live time fraction of the secondary channels.

\begin{figure}[h]
  \includegraphics[width=\columnwidth]{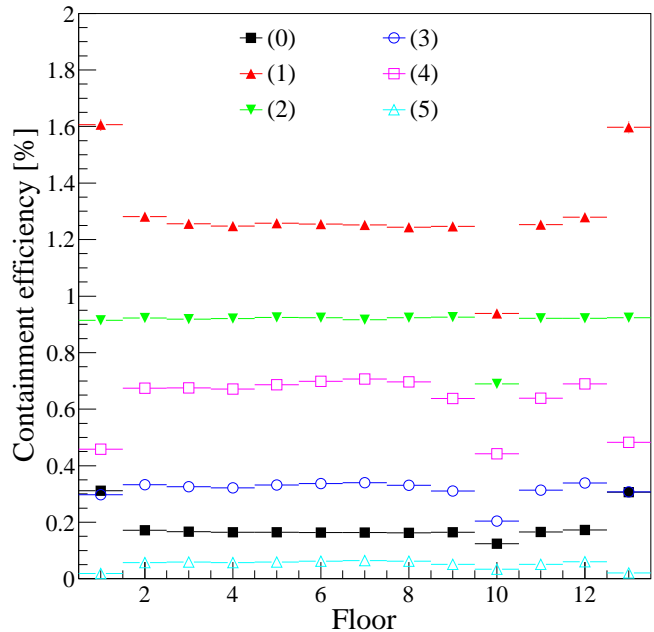}
  \caption{\label{fig:effMC}Containment efficiency $\varepsilon_{\rm MC}$ for each floor
    and for all considered \bec\ decay signatures. The uncertainty of the values
    is hidden by the markers. The efficiencies for signatures with $M>1$ are corrected
    by the live time fraction of the secondary channels.
    All efficiencies of floors 9--11 are reduced because one of the detectors
    in floor 10 was not operative.}
\end{figure}

Fig.~\ref{fig:effMC} shows $\varepsilon_{\rm MC}$ for all \bec\ decay signatures and for each single floor of \cuoreo.
As discussed in Sec.~\ref{sec:data}, the top and bottom floors feature
different efficiencies for all signatures, except signature $(2)$.
In this signature, all of the energy is deposited in a single crystal with no energy escaping,
so the efficiency is unaffected by the detector geometry (see Fig.~\ref{fig:signatures}). 
The containment efficiency for each signature and subset is also reported in Table~\ref{tab:signatures}.

\section{Statistical Approach}\label{sec:statistics}

We search for \bec\ decay by means of a simultaneous unbinned Bayesian fit
of the energy spectra of all signatures and subsets using the BAT software package~\cite{Caldwell:2008fw}.
The likelihood function is the product over the signatures (index $s$) and subsets (index $d$)
of the unbinned (extended) terms:
\begin{equation}\label{eq:extendedlikelihood}
  \mathcal{L}  = \prod_{s=0}^5 \prod_{d=0}^1  \frac{\lambda_{sd}^{n_{sd}} \cdot e^{\mbox{-}\lambda_{sd}}}{n_{sd}!}
  \prod_{i=1}^{n_{sd}}f\left( \vec{E}_{sdi} | \vec{\theta}_{sd} \right)\ ,
\end{equation}
where $n_{sd}$ is the number of events in the spectrum $sd$,
$\lambda_{sd}$ is the corresponding expectation value,
$f$ is the expected energy distribution of the signal and background events,
$\vec{E}_{sdi}$ represents the event energy values
and $\vec{\theta}_{sd}$ are the model parameters.
We drop the indexes $s$ and $d$ from here on where not necessary.

The expectation value $\lambda$ is the sum of the expected number of signal events $S$,
of linear background events $B$, and of events belonging to other possible background components.
For all signatures, we can consider the presence of additional background peaks
in the vicinity of the \bec\ decay peak position. We indicate these with an index $p$ and
a number of expected events $P_p$.
Moreover, the signatures with multiplicity $>1$ can feature the presence
of additional background events distributed on horizontal, vertical or diagonal bands,
as can be seen in Fig.~\ref{fig:Spectrum_s34}.
Namely, events in which a background $\gamma$ undergoes a Compton scattering in one crystal
and then is fully absorbed in a neighbor one, so that the two energy depositions sum up at the energy of the $\gamma$,
are distributed on a diagonal band.
On the other hand, a background $\gamma$ can undergo a pair production followed by an electron-positron annihilation.
If one of the $511$~keV $\gamma$ rays is absorbed in a neighbor crystal,
while the other undergoes a Compton scattering within the original crystal
and then escapes undetected, we measure a M2 event with $511$~keV in the neighbor crystal
and somewhat less than the energy of the SEP in the original one.
Events of this type are distributed on a horizontal band.
Finally, a background event can consist of two $\gamma$ rays emitted in coincidence:
if one is fully absorbed and the other undergoes a Compton scattering in a neighboring crystal and then escapes,
the event lies in a vertical band. This can occur in $^{60}$Co events, for example.
We will refer to the horizontal/vertical bands with an index $t$ (and expectation value $T_t$)
and to the diagonal ones with an index $q$ (and expectation value $Q_q$).
The expectation value is therefore given by:
\begin{equation}\label{eq:expectation}
  \lambda_{sd} = S_{sd} + B_{sd} + \sum_p P_{psd} + \sum_t T_{tsd} + \sum_q Q_{qsd}
\end{equation}
For the background contributions we use the number of background events
as a fit parameter, while for the signal contribution we express $S$ as a function of the \bec\ half-life:
\begin{equation}\label{eq:expectationsignal}
  S_{sd} = \frac{\ln{2}}{T_{1/2}} \cdot \frac{ N_A }{ m_A } f_{120} \cdot \varepsilon_{sd} \cdot m_d \cdot t_d\ ,
\end{equation}
where $N_A$ is the Avogadro number, $m_A$ is the molar mass of \TeOO,
$f_{120}$ is the \Tee\ isotopic abundance, $\varepsilon_{sd}$ is the total efficiency,
i.e. the product of the containment efficiency $\varepsilon_{{\rm MC},sd}$,
the coincidence efficiency for the considered signature multiplicity $\varepsilon_{\rm M}$ ($M=1,2,3$)
and the selection cut efficiency $\varepsilon_{\rm cut}$\footnote[2]{
  Notice that $\varepsilon_{{\rm MC},sd}$ is different for each signature and subset,
  $\varepsilon_{\rm Mi}$ depends on the event multiplicity for the considered signature,
  while $\varepsilon_{\rm cut}$ is common to all signatures and subsets.},
while $m_d$ and $t_d$ are the TeO$_2$ mass and the measurement live time of subset $d$, respectively.

We model the energy distribution for every signature and subset according to the contributions
considered in Eq.~(\ref{eq:expectation}). In general, we can express $f(\vec{E}|\vec{\theta})$ as:
\begin{align}\label{eq:pdf}
  f\left( \vec{E}_i | \vec{\theta} \right) = 
  & \frac{S}{\lambda} f_S \left( \vec{E}_i | \vec{\theta} \right) & \rightarrow \text{signal} \nonumber \\
  + & \frac{B}{\lambda} f_B \left( \vec{E}_i | \vec{\theta} \right) & \rightarrow \text{linear background} \nonumber \\
  + & \sum_p \frac{P_p}{\lambda} f_{p} \left( \vec{E}_i | \vec{\theta} \right) & \rightarrow \text{background peaks} \nonumber \\
  + & \sum_t \frac{T_t}{\lambda} f_{t} \left( \vec{E}_i | \vec{\theta} \right) & \rightarrow \text{hor./vert. bands} \nonumber \\
  + & \sum_q \frac{Q_q}{\lambda} f_{q} \left( \vec{E}_i | \vec{\theta} \right) & \rightarrow \text{diagonal bands.}
\end{align}
For all signatures, we parameterize the signal as an $M$-dimensional Gaussian distribution
centered at the energies $\vec{\mu}$ reported in Table~\ref{tab:signatures}:
\begin{equation}\label{eq:pdfsignal}
  f_S \left( \vec{E}_i | \vec{\theta} \right) =
  \prod_{r=1}^M \frac{1}{\sqrt{2\pi} \cdot \sigma_r}
  \exp{ \left[ - \frac{ \left(E_{ir} - \mu_{r}\right)^2 }{2 \sigma_{r}^2} \right] }\ ,
\end{equation}
where $r$ is the dimension index running from 1 to the considered multiplicity $M$.

The distribution of background events depends both on the considered signature and subset.
In general, we implement it as a linear distribution in all considered dimensions:
\begin{equation}\label{eq:pdfbackground}
  f_B \left( \vec{E}_i | \vec{\theta} \right) =
  \prod_{r=1}^M \left[ \frac{1}{\Delta E_r} + \beta \cdot ( E_{ir} - \hat{E}_r ) \right]\ ,
\end{equation}
where $\Delta E_r$ and $\hat{E}_r$ are the fit range and its center, respectively,
for the dimension index $r=1,\dots,M$,
while $\beta$ is a parameter which describes the slope of the background distribution.

In case other background peaks are present in the fit region, we parameterize them
as n-dimensional Gaussian peaks centered at energy $\vec{\mu}_{p}$
and with sigma $\vec{\sigma}_p$:
\begin{equation}\label{eq:pdfbkgpeaks}
  f_p \left( \vec{E}_i | \vec{\theta} \right) =
  \prod_{r=1}^M \frac{1}{ \sqrt{2\pi}\cdot\sigma_{pr} }
  \exp{ \left[ - \frac{ \left(E_{ir} - \mu_{pr}\right)^2 }{2\sigma_{pr}^2} \right] }\ .
\end{equation}

The signatures with multiplicity $>1$ can feature the presence of horizontal and/or vertical background bands,
which we implement as:
\begin{equation}\label{eq:pdfhorverbands}
  f_t \left( \vec{E}_i | \vec{\theta} \right) =
  \frac{1}{\sqrt{2\pi}\sigma_t} \frac{1}{\Delta E_{|1-k|}}
  \exp{ \left[ - \frac{ \left( E_{ik} - \mu_t \right)^2 }{ 2\sigma_t^2 } \right] }\ ,
\end{equation}
where $k$ is the index indicating the direction of the band in the 2-dimensional spectrum.

Finally, we fit the diagonal bands with:
\begin{equation}\label{eq:pdfdiagbands}
  f_q \left( \vec{E}_i | \vec{\theta} \right) =
  \frac{
    \exp{ \left[ - \frac{ \left( E_{i0} + E_{i1} - \mu_q \right)^2 }{ 2 \sigma_q^2 } \right] }
  }{
    \bigintsss_{\Delta E_1, \Delta E_2}
    \exp{ \left[ - \frac{ \left( E_{i0} + E_{i1} - \mu_q \right)^2 }{ 2 \sigma_q^2 } \right] }dE_1dE_2
  }
  \ ,
\end{equation}
where $E_{i0}$ and $E_{i1}$ are the energies measured in the two crystals,
$\mu_q$ is the energy of the original $\gamma$, and $\sigma_q$ is the combination
of the energy resolution in the two channels: $\sigma_q = \sqrt{\sigma_1^2+\sigma_2^2}$.

The fit parameters with no prior information available are the normalization terms for the background contributions
$B_{sd}$, $P_{psd}$, $Q_{qsd}$ and $T_{tsd}$,
and the parameter of interest is the \bec\ inverse half-life, $1/T_{1/2}$.
The nuisance parameters for which prior measurements are available
are the containment efficiencies $\varepsilon_{\rm MC}$,
the coincidence cut efficiencies $\varepsilon_{\rm M}$,
the selection cut efficiency $\varepsilon_{\rm cut}$,
the \Tee\ isotopic abundance $f_{120}$ and the \bec\ Q-value $Q$.
The last three parameters and the inverse half-life are common to all signatures.

We use two sets of priors for the fit parameters. If an independent measurement
is available for a parameter, we use a Gaussian prior centered at the measured value
and with a $\sigma$ equal to the corresponding uncertainty.
This is the case for the efficiencies, $Q$ and $f_{120}$.
For all other parameters we use a flat prior in a range large enough to allow
the corresponding marginalized posterior to go to zero,
and bound to non-negative values if the considered parameter represents
or is proportional to a number of counts.
The choice of a flat prior does not influence significantly the posterior
for the background components, because the information contained in the data
is generally stronger than that provided by the prior.
This is not true for the very small or negligible background components -- with a posterior peaked
at zero or compatible at $2~\sigma$ with it -- or $1/T_{1/2}$.
In these cases, a log-flat prior, i.e. a prior flat in the logarithm of the variable,
would yield a much stronger limit on the number of counts
assigned to the considered background components, or on $1/T_{1/2}$.
Therefore, the flat prior represents a conservative choice.

The inclusion of the energy resolution $\sigma$ as a nuisance parameter
would involve a further complication of the analysis software.
Namely, while all other parameters can be either characteristic of each single fit component,
or common to all of them (within the same signature and subset),
the energy resolution is common to multiple background components of different signatures,
and at the same time different background components of the same signature can have different resolutions.
For sake of simplicity, we preferred to treat the energy resolution as a systematic effect
and to run the analysis multiple times after shifting all $\sigma$ values up and down
by their uncertainties.

\section{Results}\label{sec:results}

\begin{figure*}[h]
  \includegraphics[width=\columnwidth]{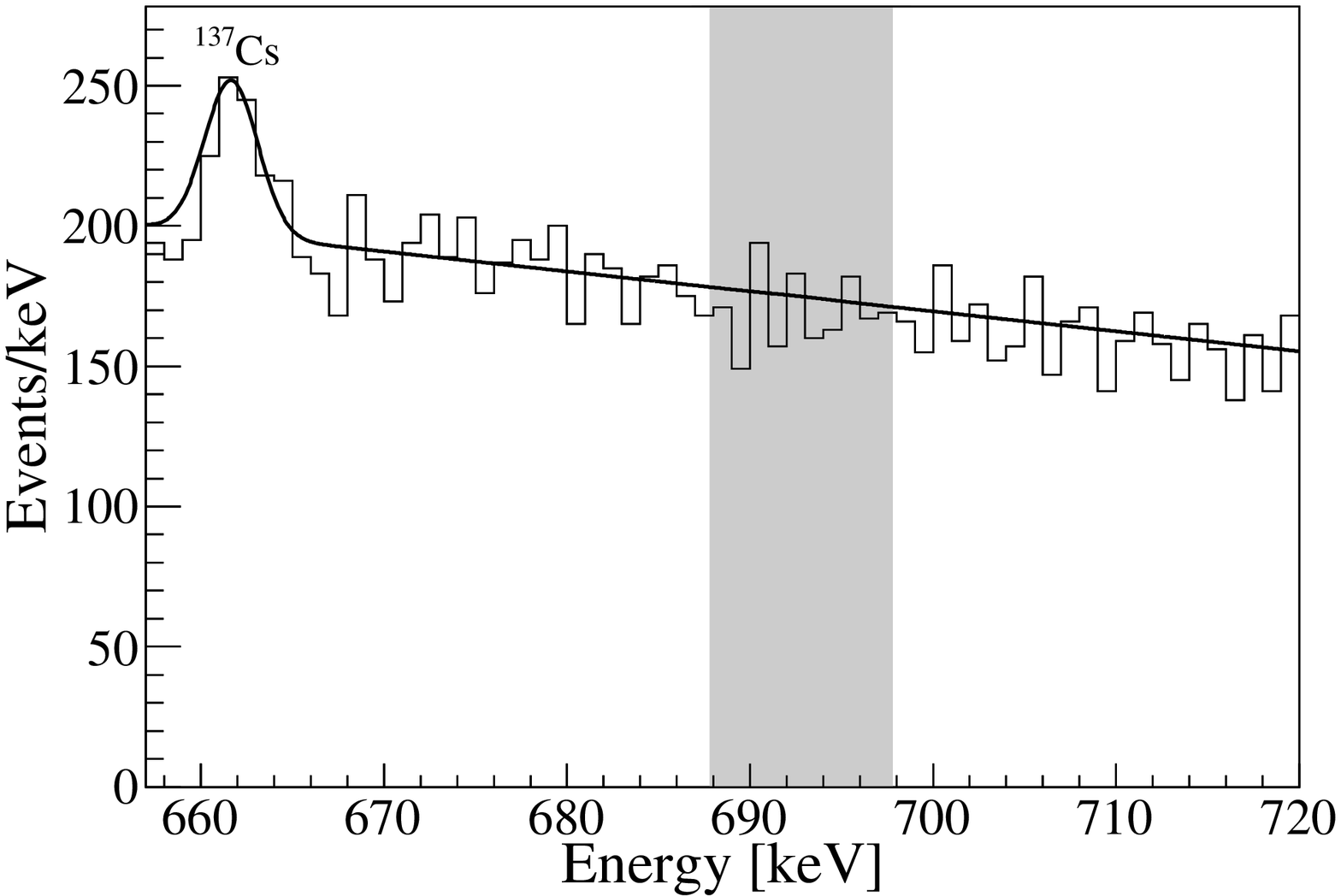}\quad
  \includegraphics[width=\columnwidth]{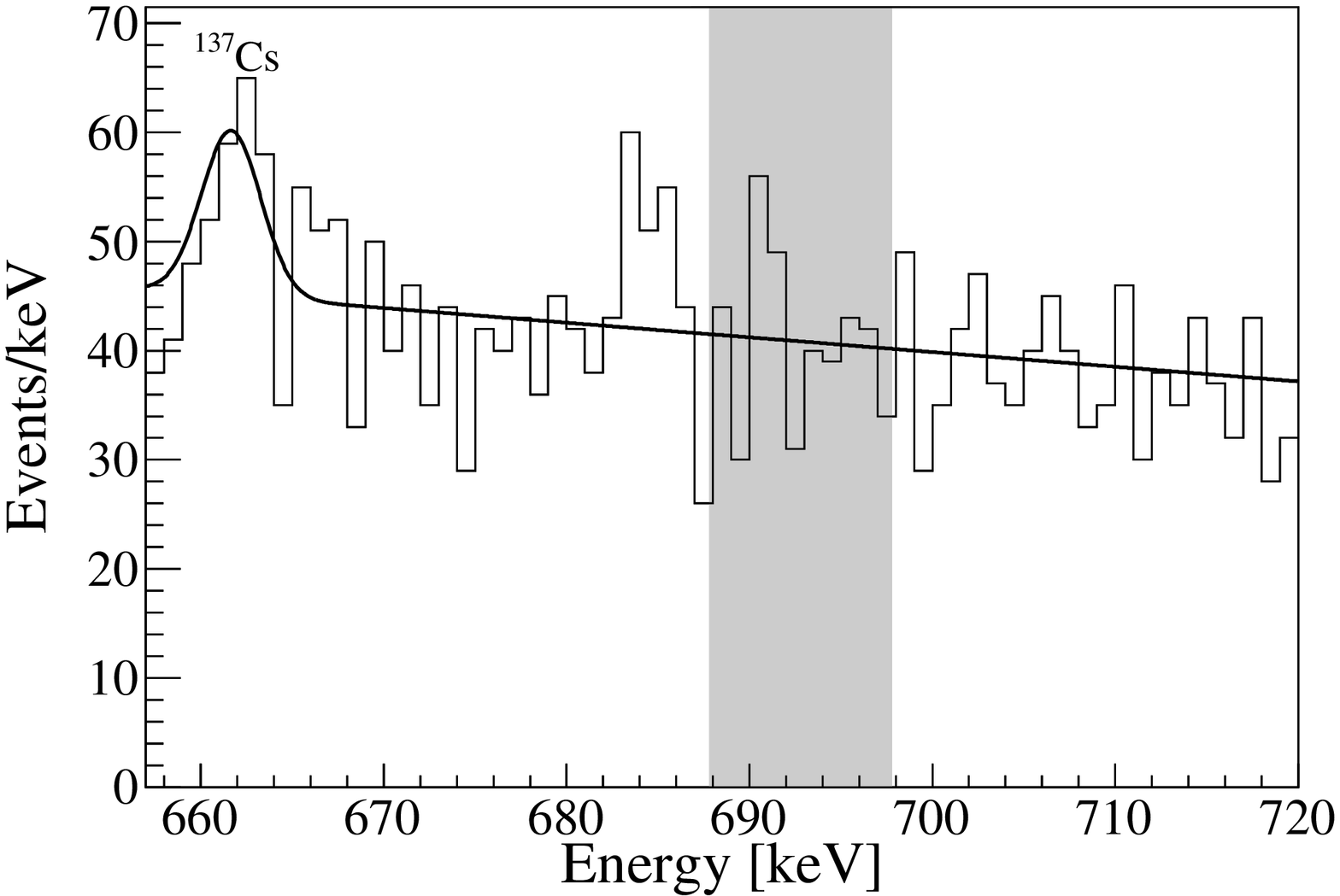}\\
  \ \\
  \includegraphics[width=\columnwidth]{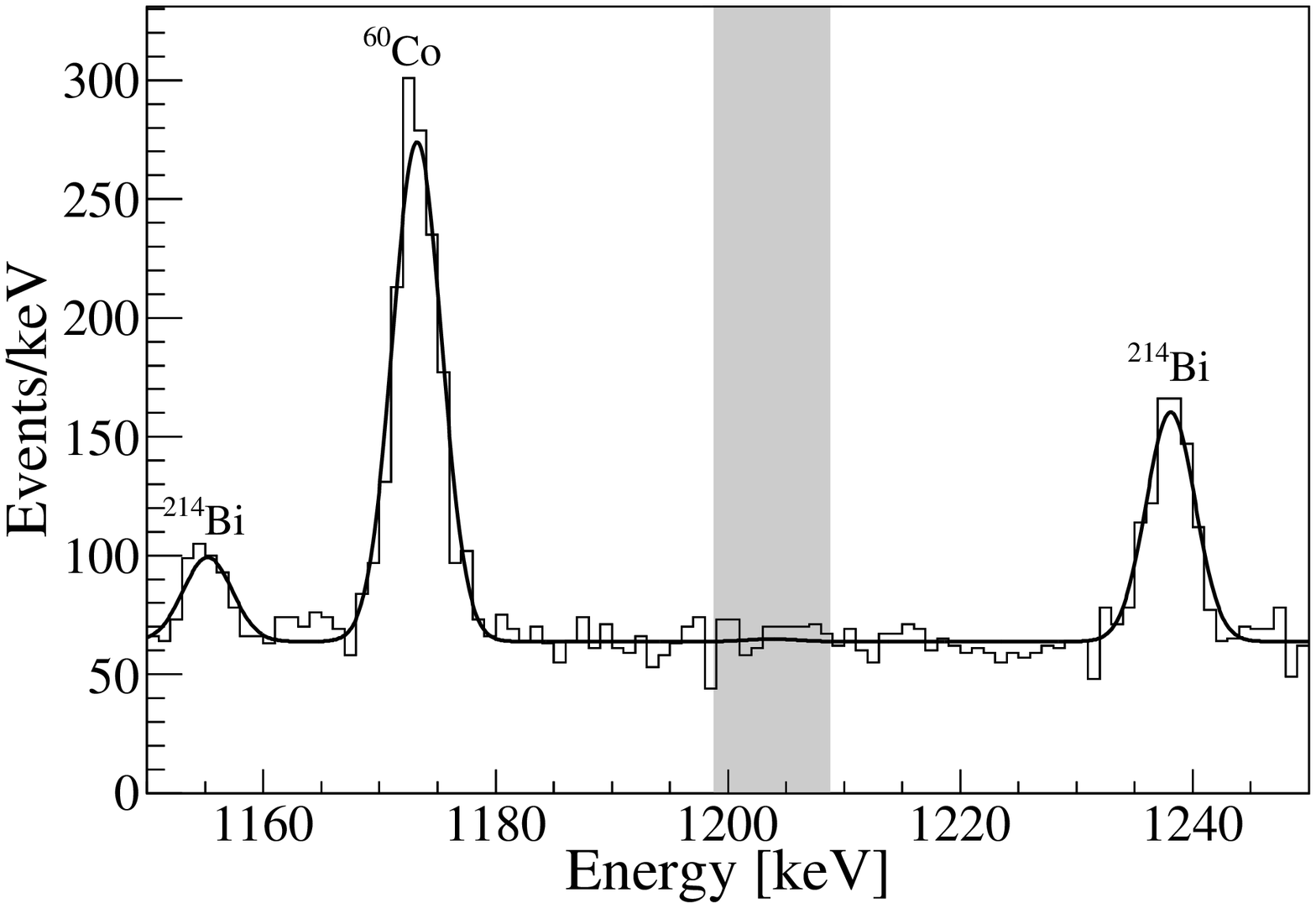}\quad
  \includegraphics[width=\columnwidth]{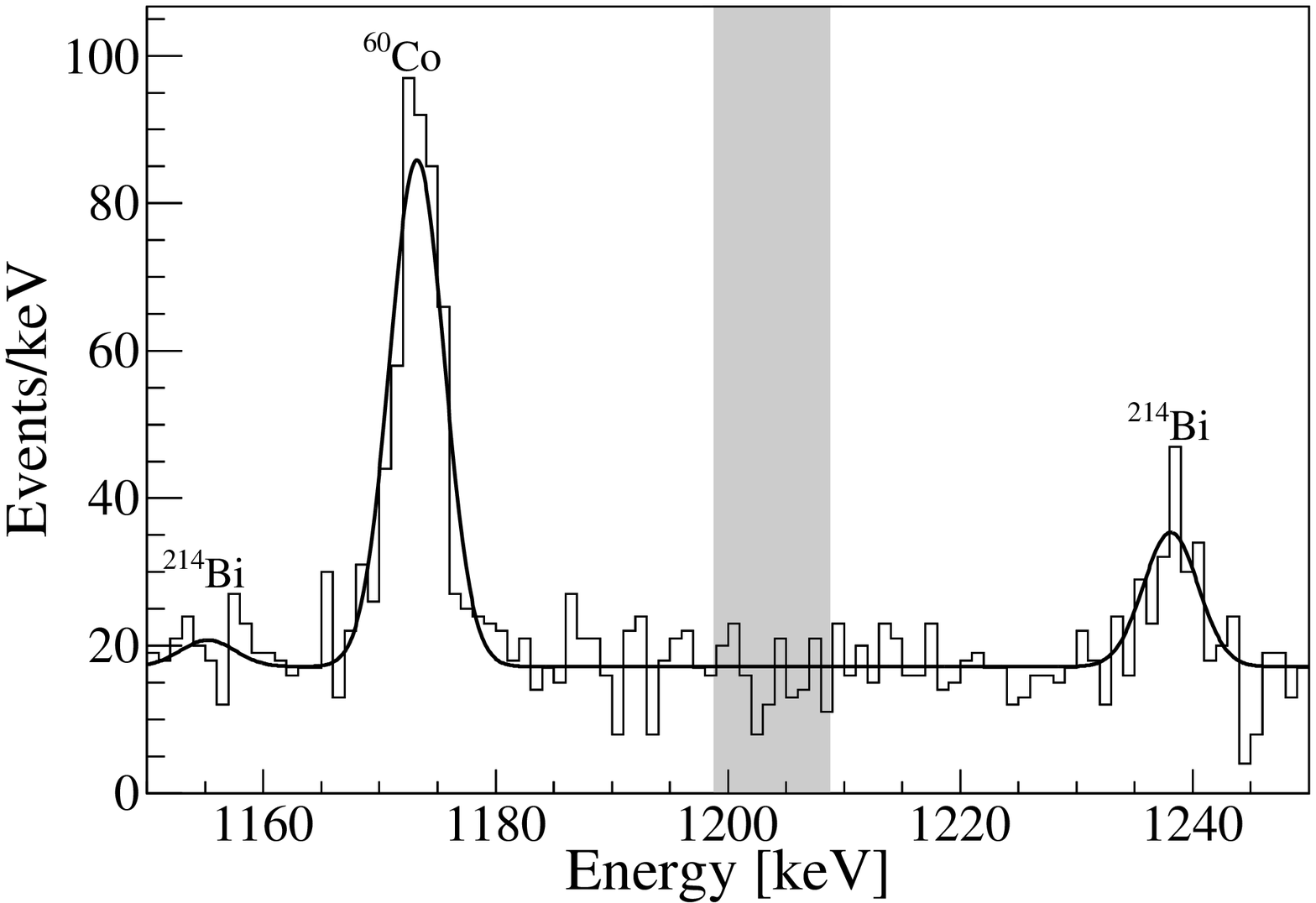}\\
  \ \\
  \includegraphics[width=\columnwidth]{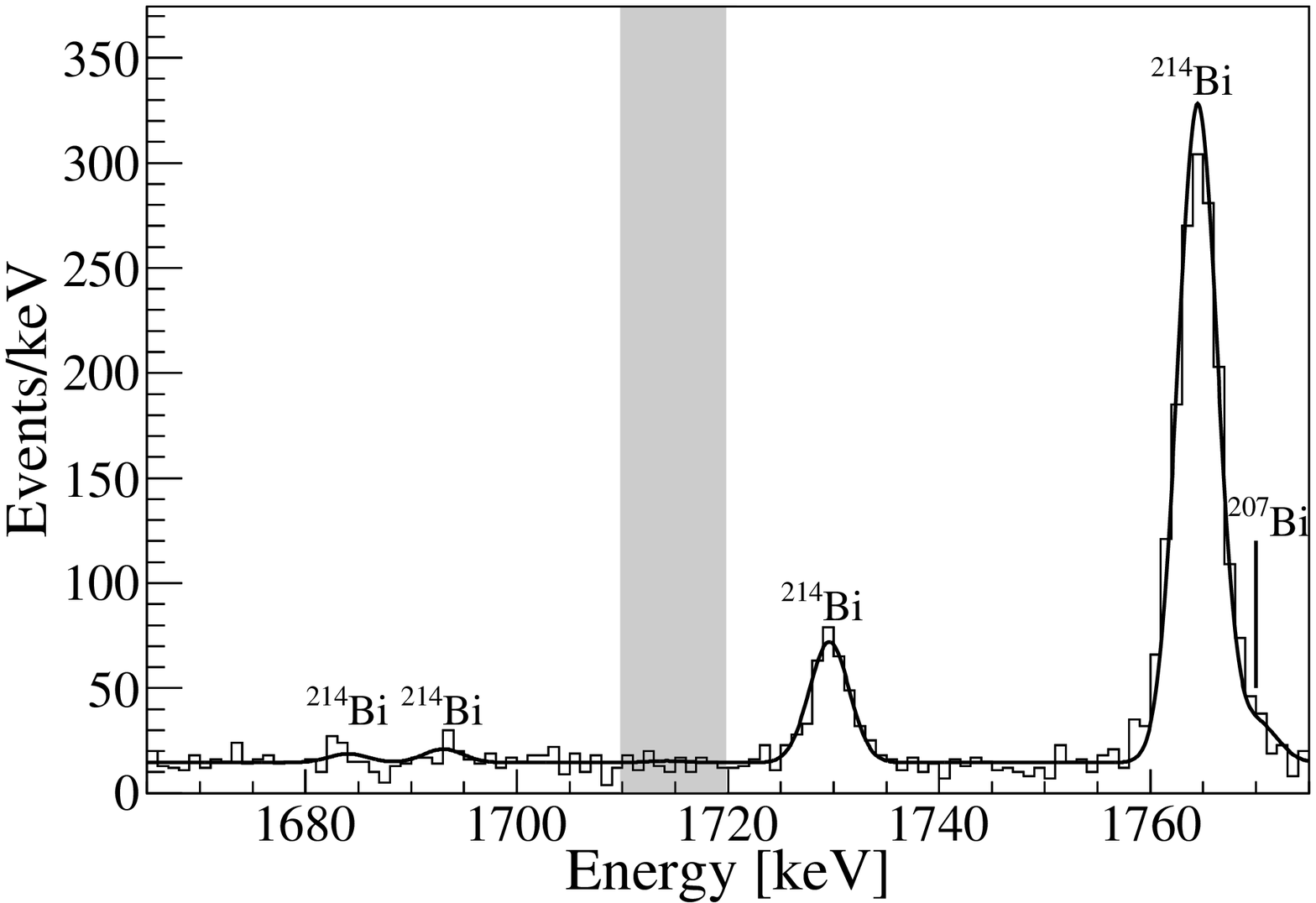}\quad
  \includegraphics[width=\columnwidth]{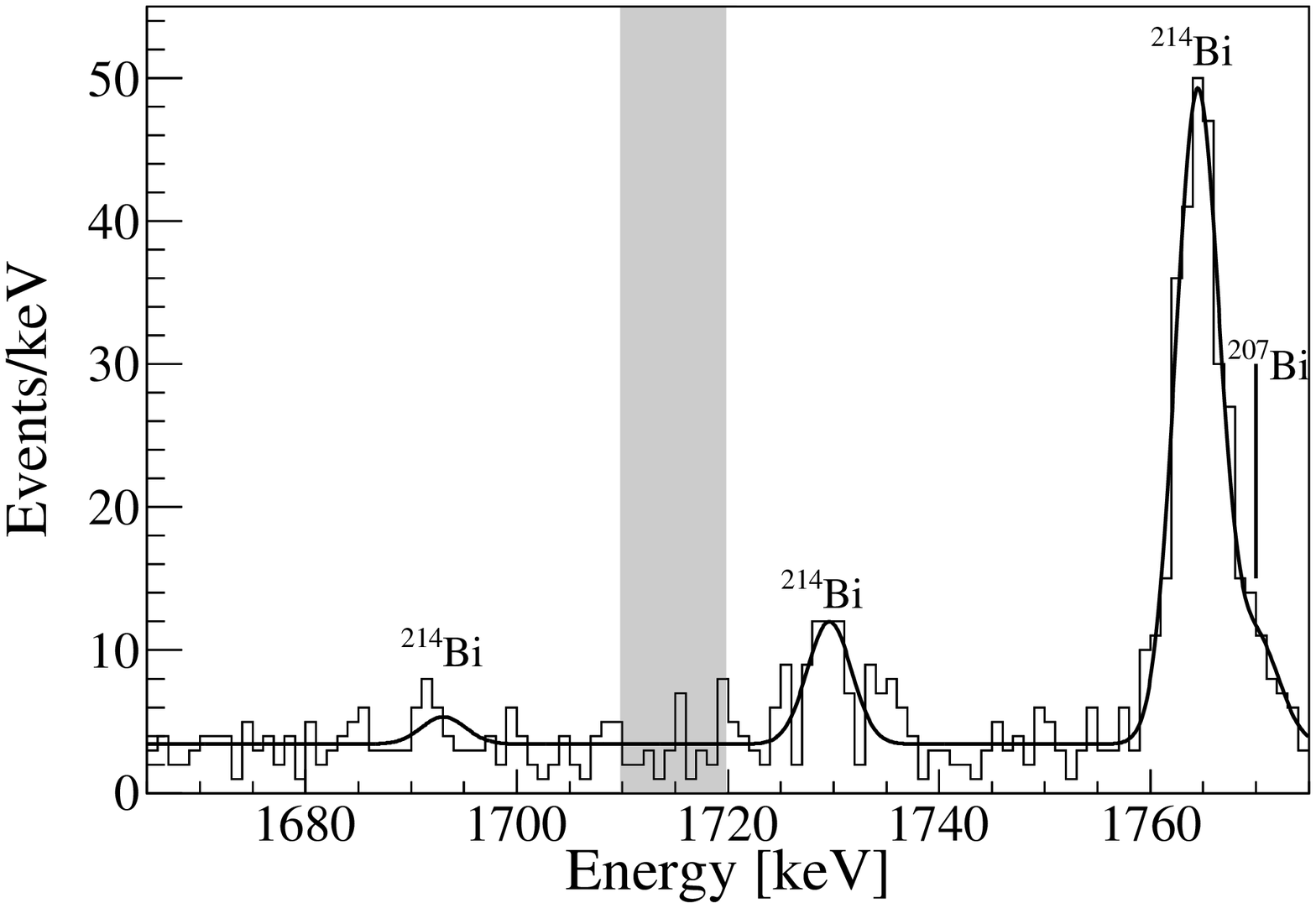}
  \caption{\label{fig:Spectrum_s012}Energy spectra for floors 2--12 (left) and 1,13 (right)
    relative to the signatures (0), (1) and (2) (top to bottom).
    The curves correspond to the best fit minimal model, with the \bec\ decay peak
    normalized to the $90\%$~CI limit. The shaded area corresponds to a 10~keV region
    around the expected signal peak position.}
\end{figure*}

\begin{figure*}[h]
  \includegraphics[width=\columnwidth]{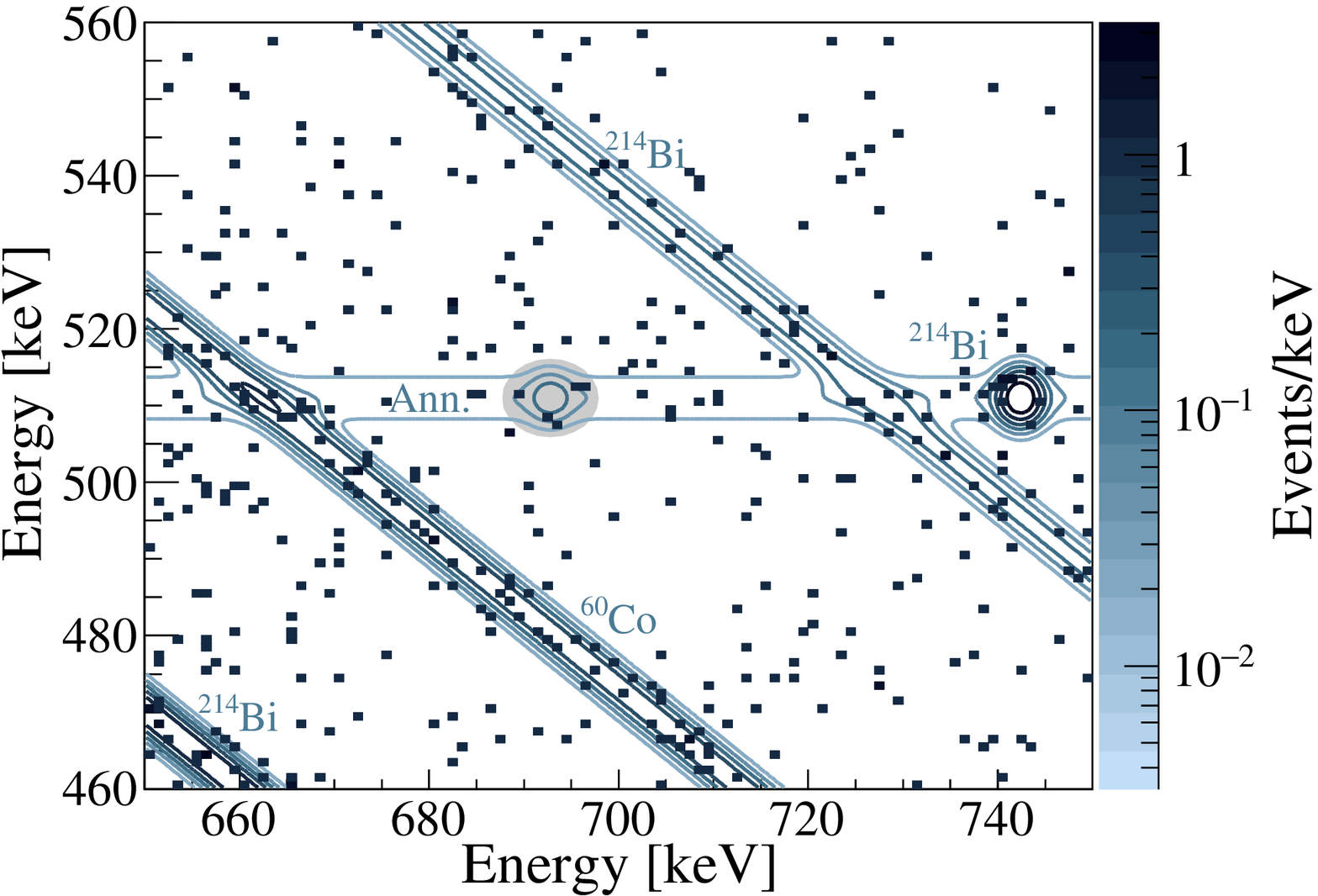}\quad
  \includegraphics[width=\columnwidth]{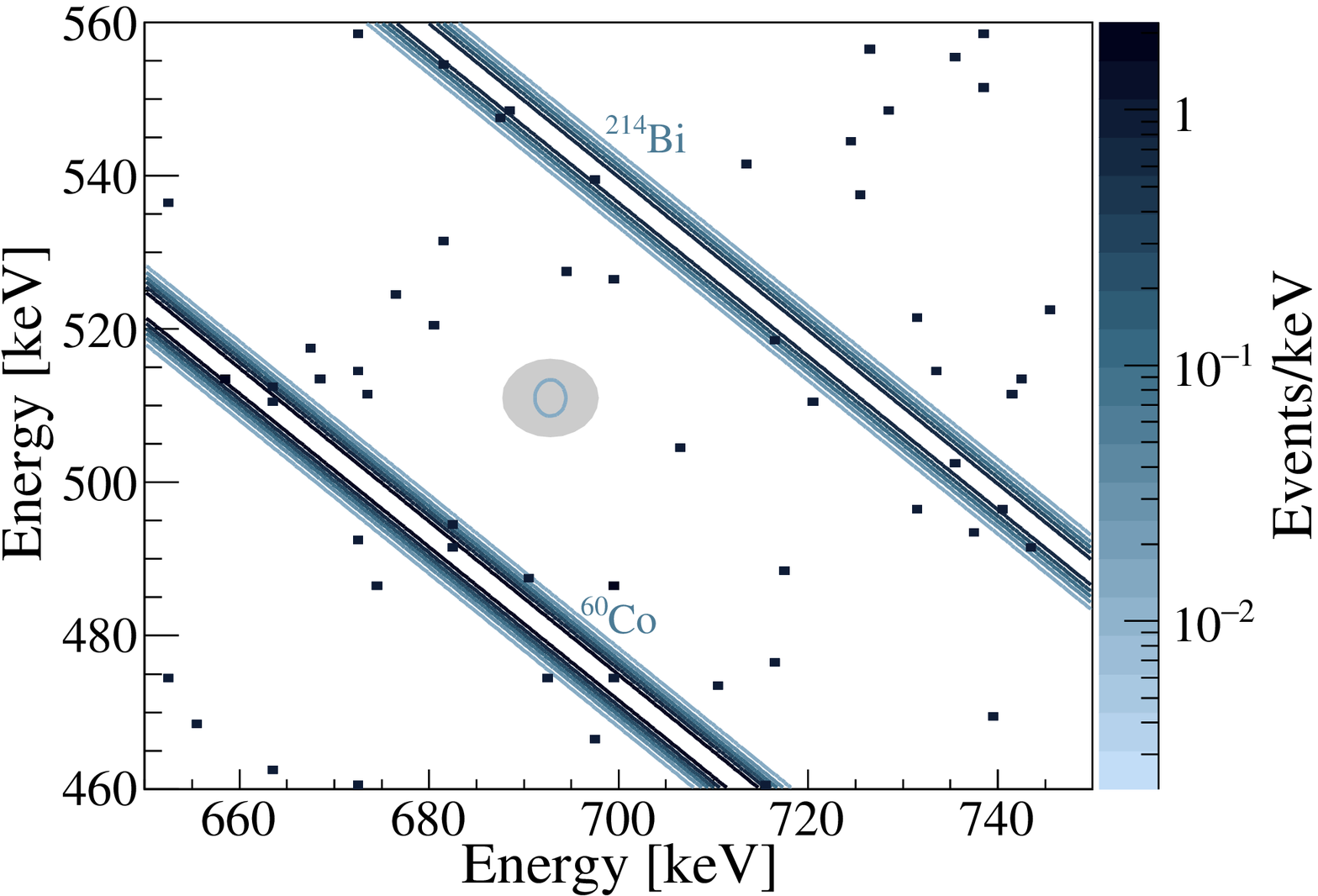}\\
  \ \\
  \includegraphics[width=\columnwidth]{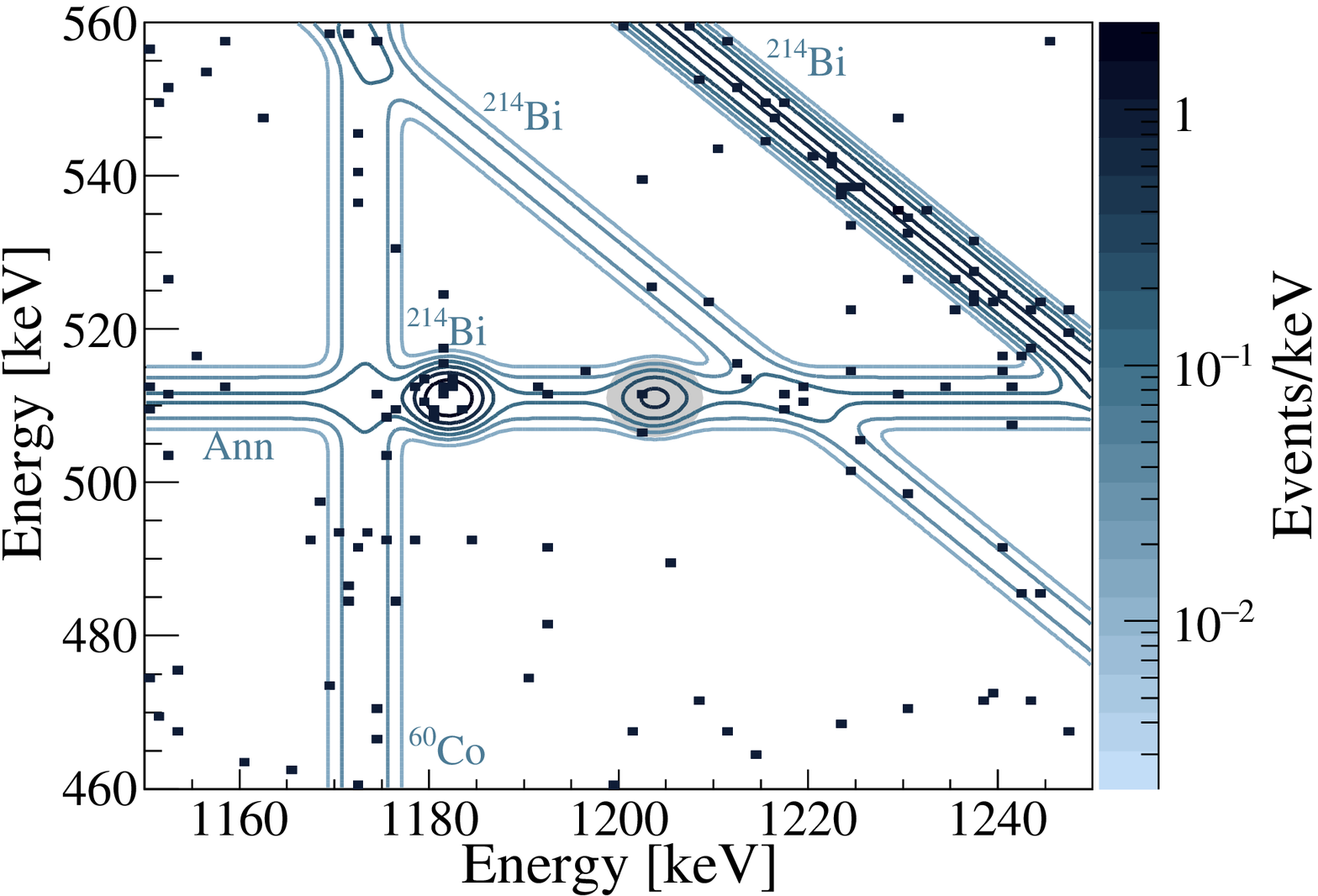}\quad
  \includegraphics[width=\columnwidth]{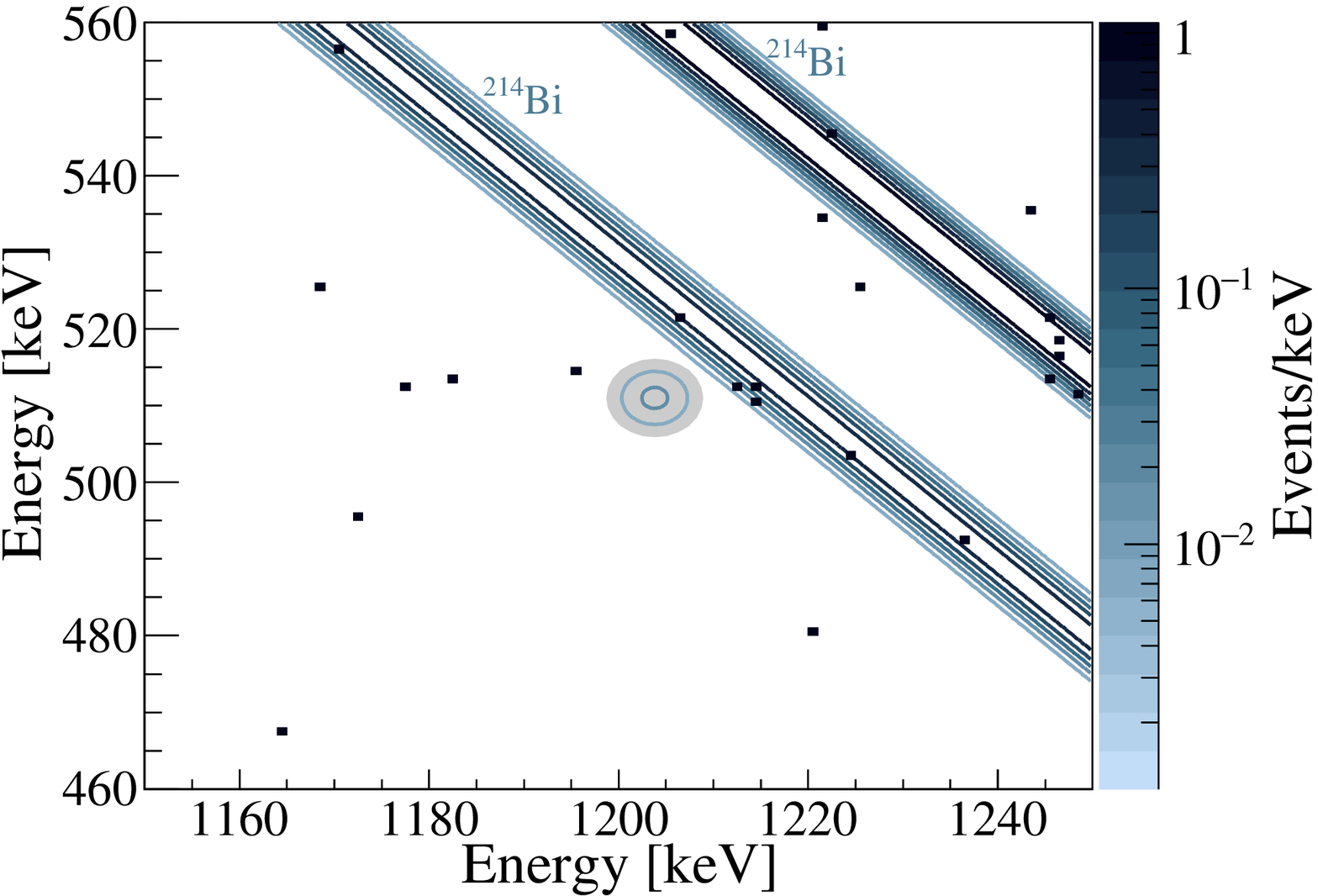}
  \caption{\label{fig:Spectrum_s34}Energy spectra for floors 2--12 (left) and 1,13 (right)
    relative to the signatures (3) and (4) (top to bottom).
    The squares correspond to the measured events.
    The contour curves correspond to the best fit minimal model,
    with the \bec\ decay contribution normalized to the $90\%$~CI limit.
    The shaded area corresponds to a 10~keV region around the expected signal peak position.}
\end{figure*}

We use the MC simulations~\cite{Alduino:2016vtd} described in Sec.~\ref{sec:mc}
to define a maximal model containing all possible background contributions.
We fit the maximal model to the data and iteratively remove those contributions
for which the minimum of the $95\%$ interval around the marginalized mode is zero.
The only exception is made for the linear background contribution, which we always keep in the fit.
We denote the final fit model containing only the components with $>2\,\sigma$ significance the ``minimal model''.
To better understand the relative importance of each signature
and the effect of the nuisance parameters, we perform the fit
on each signature separately, as well as on all of them together,
and under the following conditions:
\begin{itemize}[noitemsep,nolistsep,leftmargin=*]
\item with the minimal model and including the efficiencies, $Q$ and $f_{120}$
  as nuisance parameters. This is our baseline approach;
\item with the minimal model, keeping the efficiencies, $Q$ and $f_{120}$ fixed;
\item with the maximal model and including the efficiencies, $Q$ and $f_{120}$
  as nuisance parameters. 
\end{itemize}
The background components of the minimal model for signatures (0)--(4)
are shown in Figs.~\ref{fig:Spectrum_s012} and~\ref{fig:Spectrum_s34}.
These figures also show the best fit curves with the \bec\ decay signal contribution
normalized to the $90\%$~credibility interval (CI) limit.
Signature (5) has only 8 events hence we parameterize its background with a uniform distribution.

\begin{table}[h]
  \caption{Results of the \bec\ decay analysis
    on each individual signature, as well as on the combination of all signatures.
    We use the minimal model keeping the efficiencies, $Q$ and $f_{120}$ fixed (column ``Less Pars.''),
    or considering them as nuisance parameters (column ``All Pars.'').
    We consider the maximal model only in the case of signatures (2), (3), (4), and for the combination of all signatures.
    For signatures (0), (1) and (5) the maximal model is equivalent to the minimal model with additional nuisance parameters.
  }
  \label{tab:results}
  \begin{ruledtabular}
    \begin{tabular}{lccc}
      & \multicolumn{3}{c}{Limit on $T_{1/2}$~[yr]}\\
      & \multicolumn{2}{c}{Minimal} & Maximal \\
      Signature & Less Pars.   & All Pars.    & All Pars. \\
      \hline\\ [-2ex]
      (0) & $2.8\cdot10^{19}$ & $2.5\cdot10^{19}$ & - \\
      (1) & $1.6\cdot10^{20}$ & $1.4\cdot10^{20}$ & - \\
      (2) & $4.7\cdot10^{20}$ & $4.2\cdot10^{20}$ & $4.2\cdot10^{20}$ \\
      (3) & $5.2\cdot10^{20}$ & $4.4\cdot10^{20}$ & $4.4\cdot10^{20}$ \\
      (4) & $1.2\cdot10^{21}$ & $1.1\cdot10^{21}$ & $1.1\cdot10^{21}$ \\
      (5) & $1.6\cdot10^{20}$ & $1.5\cdot10^{20}$ & - \\
      \hline\\ [-2ex]
      All & $1.6\cdot10^{21}$ & $1.6\cdot10^{21}$ & $1.6\cdot10^{21}$ \\
    \end{tabular}
  \end{ruledtabular}
\end{table}

The results of all fits are reported in Table~\ref{tab:results}.
The inclusion of additional nuisance parameters for the efficiencies, $Q$ and $f_{120}$
weakens all limits by $\sim10\%$, with the largest effect obtained for signature (2)
and the greatest reduction coming from the uncertainty on $f_{120}$.
The effect is not the same for all subsets:
the reason comes from the presence of additional statistical fluctuations
to which the fit becomes sensitive when $Q$ is not constrained to its best fit value.
When we run the fits simultaneously on all signatures, the inclusion of additional nuisance parameters
affects the limit by just $1\%$.
Additionally, the effect of the switch to the maximal model is at the percent level for all signatures,
and indicates that the minimal model already provides an appropriate description of the data.

Finally, we consider the uncertainty on the energy resolution as a systematic.
We re-run the minimal model on the data with all efficiencies, $Q$ and $f_{120}$ as nuisance parameters
and with the energy resolution increased or decreased by $\pm1$ standard deviation.
This variation yields a $\mp7\%$ change in the $T_{1/2}$ limit, respectively.
With the described procedure we are neglecting all correlations between the uncertainties
reported in Table~\ref{tab:fwhm}. Hence, this result represents a conservative estimation.

The fit of the minimal model on all signatures together with the inclusion of the efficiencies, $Q$ and $f_{120}$
as nuisance parameters, and without considering the systematic induced by energy resolution, gives a limit of
\begin{equation}
  T_{1/2} > 1.6\cdot10^{21}~\text{yr} 
\end{equation}
for a $90\%$~CI.
The limit obtained with \cuoreo\ data is slightly weaker than that achieved with \cuoricino.
Specifically, the limit is weakened by the presence of a small upward fluctuation
in signature (1), with significance of $\sim1~\sigma$.

\begin{figure}[h]
  \includegraphics[width=\columnwidth]{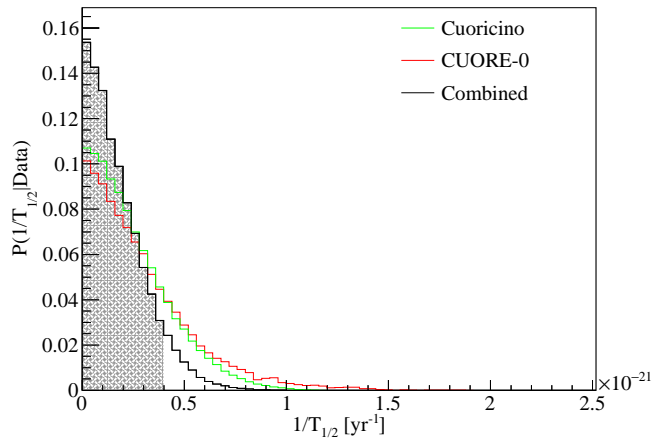}
  \caption{Posterior distribution of $1/T_{1/2}$ obtained from \cuoricino\ (green),
    CUORE-0 (red), and their combination (black). The gray area corresponds to the $90\%$
    quantile.}
  \label{fig:posteriors}
\end{figure}

We combine the results of \cuoreo\ and \cuoricino\ through a Bayesian fit with a flat prior on $1/T_{1/2}$,
rescaling the \cuoricino\ result to account for the corrected isotopic abundance.
Thus, we obtain the strongest limit to date on \bec\  decay of \Tee:
\begin{equation}
  T_{1/2} > 2.7\cdot10^{21}~\text{yr (}90\%\text{~CI),}
\end{equation}
with a $5\%$ systematic uncertainty induced by the uncertainty on the \cuoreo\ energy resolution.
The $1/T_{1/2}$ posterior distribution for the combination of \cuoreo\ and \cuoricino\ results
is shown in Fig.~\ref{fig:posteriors}.

\section{Conclusions}\label{sec:conclusion}

We performed a search of \bec\ decay of \Tee\ on \cuoreo\ data.
The lower background, the higher total efficiency and the development of a dedicated fit which we can run simultaneously
on multiple signatures with different multiplicities and on multiple data subsets
allows us to reach a limit comparable to that of \cuoricino\
with approximately half the exposure.
We can apply the analysis procedure developed for this work directly to \cuore\ data, once available,
for the search of \Tee\ \bec\ decay and other physics processes,
e.g. the \onbb\ decay of \Te\ to excited states of $^{130}$Xe,
already studied with \cuoricino~\cite{Andreotti:2011in} and \cuoreo~\cite{spozzi}.
The larger mass and higher containment efficiency for events with multiplicity $>1$
could provide an increase of two orders of magnitude for all analyses of this kind.

\begin{acknowledgments}
  The \cuore\ Collaboration thanks the directors and staff of the Laboratori Nazionali del Gran Sasso
  and the technical staff of our laboratories. This work was supported by
  the Istituto Nazionale di Fisica Nucleare (INFN);
  the National Science Foundation under Grant Nos. NSF-PHY-0605119, NSF-PHY-0500337, NSF-PHY-0855314, NSF-PHY-0902171,
  NSF-PHY-0969852, NSF-PHY-1307204, NSF-PHY-1314881, NSF-PHY-1401832, and NSF-PHY-1404205; the Alfred P. Sloan Foundation;
  the University of Wisconsin Foundation; and Yale University.
  This material is also based upon work supported by the US Department of Energy (DOE) Office of Science
  under Contract Nos. DE-AC02-05CH11231, DE-AC52-07NA27344, and DE-SC0012654;
  and by the DOE Office of Science, Office of Nuclear Physics under Contract Nos. DE-FG02-08ER41551 and DE-FG03-00ER41138.
  This research used resources of the National Energy Research Scientific Computing Center (NERSC).
\end{acknowledgments}

\bibliography{Bibliography}

\end{document}